\documentclass[twocolumn,showpacs,amsmath,amssymb]{revtex4}
\usepackage{graphicx}% Include figure files
\usepackage{dcolumn}% Align table columns on decimal point
\usepackage{bm}% bold math
\usepackage{psfig}% bold math
\usepackage{slashbox}
\newcommand{\psp}{\psi^{\prime}}

\newcommand{\jpsi}{J/\psi}
\newcommand{\EE}{e^+e^-}

\newcommand{\pp}{\pi^+\pi^-}

\newcommand{\beq}{\begin{equation}}
\newcommand{\eeq}{\end{equation}}
\newcommand{\beqn}{\begin{eqnarray}}
\newcommand{\eeqn}{\end{eqnarray}}
\newcommand{\beqns}{\begin{eqnarray*}}
\newcommand{\eeqns}{\end{eqnarray*}}
\newcommand{\bfg}{\begin{figure}}
\newcommand{\efg}{\end{figure}}
\newcommand{\bitm}{\begin{itemize}}
\newcommand{\eitm}{\end{itemize}}
\newcommand{\bnum}{\begin{enumerate}}
\newcommand{\enum}{\end{enumerate}}
\newcommand{\btbl}{\begin{table}}
\newcommand{\etbl}{\end{table}}
\newcommand{\btbu}{\begin{tabular}}
\newcommand{\etbu}{\end{tabular}}
\def\eref#1{(\ref{#1})}
\def\Journal#1#2#3#4{{#1} {\bf #2}, #3 (#4)}
\def\PLB{Phys. Lett. B}

\def\PRL{Phys. Rev. Lett.}
\def\PRD{Phys. Rev. D}

\def\prd#1#2#3 {{~Phys. Rev. D {#1}, #2 (#3) }} %#1=volume, #2=page, #3=year
\def\plb#1#2#3 {{~Phys. Lett. B {#1}, #2 (#3) }} %#1=volume, #2=page, #3=year

\begin{document}
\preprint{Draft-v1}

\title{On the leptonic partial widths of the excited
$\psi$ states}
\author{X.~H.~Mo}
 \email{moxh@ihep.ac.cn}
\author{C.~Z.~Yuan}
 \email{yuancz@ihep.ac.cn}
\author{P.~Wang}
 \email{wangp@ihep.ac.cn}
\affiliation{Institute of High Energy Physics, Chinese Academy of
Sciences, Beijing 100049, China}

\date{\today}

\begin{abstract}

The resonance parameters of the excited $\psi$-family resonances,
namely the $\psi(4040)$, $\psi(4160)$, and $\psi(4415)$, were
determined by fitting the $R$-values measured by experiments. It
is found that the previously reported leptonic partial widths of
these states were merely one possible solutions among a four-fold
ambiguity. By fitting the most precise experimental data on the
$R$-values measured by the BES collaboration, this work presents
all four sets of solutions. These results may affect the
interpretation of the charmonium and charmonium-like states above
4~GeV/$c^2$.

\end{abstract}

\pacs{12.38.Qk, 13.66.Bc, 14.40.Gx}

\maketitle

\section{Introduction}

Charmonium, bound state of a charm and an anti-charm quarks, is
one of the most interesting two-body systems which were studied
extensively in particle physics. Although the first charmonium
state was discovered more than thirty-five years ago, there are still many
puzzles in charmonium physics. The charmonium spectroscopy below
the open charm threshold has been well measured and agrees with
the theoretical expectations (such as potential models and lattice
QCD); however, for the charmonium states above the
open charm threshold, there are still lack of adequate
experimental information and solid theoretical inductions. For
example, recently many new particles have been discovered, named
$XYZ$-particles, and the overwhelming vector states in the
4~GeV/$c^2$ to 5~GeV/$c^2$ mass range make the classification of
these states as the charmonia
questionable~\cite{Olsen,Swanson,QWG}. In explaining these vector
charmonium states, the leptonic partial widths
provide very important information. As we know, the
vector quarkonium states could be either $S$-wave or $D$-wave
spin-triplet states, with the $S$-wave states couple strongly to
lepton pair while the $D$-wave states couple weakly since the
latter are only proportional to the second derivative of the
wave-function at the origin squared, as expected in the potential
models. This leads people believe that the $\psi(4040)$ is the
$3S$ charmonium state, $\psi(4160)$ the $2D$ state, and
$\psi(4415)$ the $4S$ state. This has been a well accepted picture
for more than two decades before the discovery of the so-called
$Y$ particles, namely, the $Y(4008)$ and $Y(4260)$ observed in
$\EE\to \pp\jpsi$ final state~\cite{y4260_babar,y4260_belle}, and
the $Y(4360)$ and $Y(4660)$ observed in $\EE\to \pp\psp$ final
state~\cite{y4360_belle,y4360_babar}. With seven states observed
between 4.0~GeV/$c^2$ and 4.7~GeV/$c^2$, some people started to
categorize some of these as non-conventional quarkonium states,
while others tried to accommodate all of them in modified
potential models. Many of the theoretical models use the leptonic
partial widths of these states to distinguish them between $S$-
and $D$-wave assumptions~\cite{Klempt,KTChao}, and most of the
time, the values on the leptonic partial widths are cited from the
PDG~\cite{PDG} directly. Although the resonance parameters
of these excited $\psi$ states have been measured by many
experimental groups, all of them
were obtained by fitting the $R$-values measured in the relevant energy
region. The most recent ones, which were from a sophisticated fit
to the most precise $R$-values measured by the BES
collaboration~\cite{besr1,besr2}, are the only source of the
leptonic partial widths of these three $\psi$ states now quoted by
the PDG~\cite{PDG}.

In fitting to the BES data, unlike the previous analyses, the BES
collaboration considered the interference between the three resonances
decaying into the same final modes, and introduced a free
relative phase for the amplitude of each resonance~\cite{bespsires}.
The new parametrization of the hadronic cross section results in a
prononce increase of the $\psi(4160)$ mass, and significant
decrease of the leptonic partial widths of $\psi(4160)$ and
$\psi(4415)$.

As has been pointed out in a recent study~\cite{Yuan:2009gd},
there are multiple solutions in fitting one dimensional
distribution with the coherent sum of several amplitudes and free
relative phase between them. Exactly the same kind of
fit to the $R$-value is performed in the present study, multiple
solutions are indeed found in extracting the resonance parameters
of the excited $\psi$ states. The only difference between these
multiple solutions is the coupling to the $\EE$, namely the
leptonic partial width, while the masses and the widths of the
resonances remain the same for all the solutions.

In the following, we firstly introduce a simplified fit scheme
similar to that used by the BES collaboration, and extract
the resonance parameters. Then we study the multiple solution
problem in the light of the toy simulated data for
illustrative purpose. We will discuss the consequence of the
multiple solutions in fitting the $R$-values at the end of the
paper.

\section{The fit to the $R$ data}

To faciliate our study, only the $R$-values provided by the BES
collaboration are used~\cite{besr1,besr2}, and in the data
fitting, only statistical uncertainties are considered, as the
systematic errors at all the energy points are highly correlated.

We fit the $\EE$ annihilation cross section $\sigma (\EE)=R \cdot
86.85/s$ ($s$ in $\hbox{GeV}^2$ and $\sigma$ in nb). The standard
chisquare estimator is constructed as
 \beq
 \chi^2 = \sum\limits_{l=1}^{N_{pt}}
 \frac{(\sigma^{exp.}_l - \sigma^{the.}_l )^2}{ (\Delta
 \sigma^{exp.}_l)^2}~,
 \label{chsqfm}
 \eeq
where $\sigma^{exp.}_l$ and $\Delta \sigma^{exp.}_l $ indicate
respectively the experimentally measured cross section and its
error at the $l$-th energy point (the number of points is denoted
as $N_{pt}$), while $\sigma^{the.}_l$ is the corresponding
theoretical expectation, which is composed of two parts
 \beq
 \sigma^{the.}(s)  = \sigma^{res.}(s)  + \sigma^{con.}(s) ~.
 \label{sigfm}
 \eeq
Here $\sigma^{con.}$ is the contribution from continuum and is
parameterized simply as
 \beq
 \sigma^{con.} (s)  = A+B (\sqrt{s}-2M_{D^{\pm}})~,
 \label{sigbg}
 \eeq
where $A$ and $B$ are free parameters, and $M_{D^{\pm}}$ is the
mass of the charged $D$ meson. $\sigma^{res.}$ is the contribution
from the resonances. Here following previous
analyses~\cite{sethr,bespsires}, the assumption that the continuum
production and the resonance decays don't interfere with each
other is adopted.
The three wide resonances shown in the data are close and have
the same decay modes, the interferences between them must be
included. Then the amplitude reads
 \beq
 T_j(s) = \frac{\sqrt{ 12 \pi \Gamma_j^h \Gamma_j^{ee} }\, e^{i \phi_j}}
 { s -M_j^2 + i M_j \Gamma_j^t }~,
 \label{ampfm}
 \eeq
with $\Gamma_j^t$, $\Gamma_j^h$, $\Gamma_j^{ee}$, and $M_j$
denoting total width, partial width to hadrons, partial width to
$\EE$ pair, and mass of the resonance $j$, respectively. The total
amplitude, which is the coherent sum of the three resonances, once
squared, contains the interferences of the type $\Re\, T_i^* T_j$.
Here $\phi_j$ is a phase associated to resonance $j$. If the
resonances are quite broad, the interference effect will distort
the shape of the resonances, the width might appear broader or
narrower, and the position of the peak might be shifted as well.

The total cross section of the resonances
 \beq
 \sigma^{res.} (s) = \left|\sum\limits_{j=1}^{3} T_j(s)
 \right|^2~.
 \label{sum_phs}
 \eeq
Since what we actually obtain is the squared module of the
amplitude, only two relative phases are relevant.

\begin{figure}[htbp]
\includegraphics[width=8.0cm]{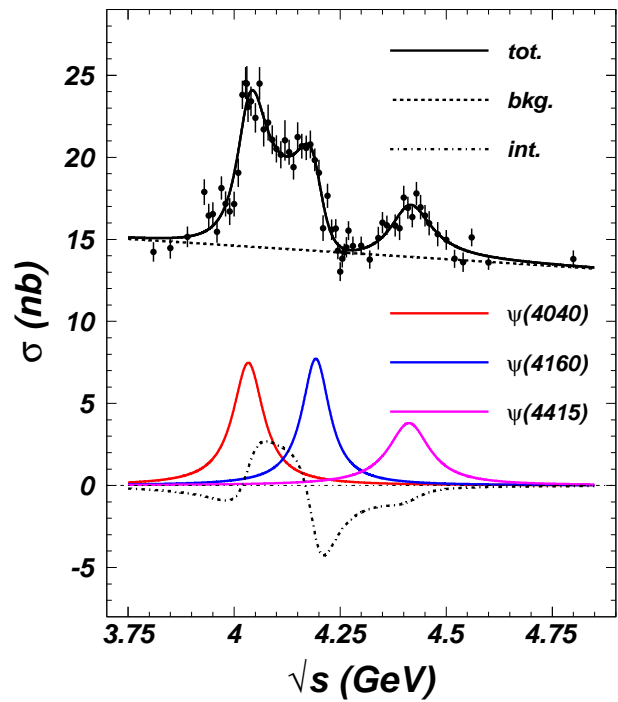}
\caption{\label{fig_datfit}Total hadronic cross section in nb
obtained as $\sigma (\EE \to \mbox{hadrons}) = R \cdot 86.85/s$
($s$ in GeV$^2$) from $R$ measurements of BES~\cite{besr1,besr2}
and the fit to data. The curves show the best fit (identical for
all the four solutions), the contribution from each resonance, as
well as the interference term (one of the four solutions).}
\end{figure}

By minimizing the $\chi^2$ defined in Eq.\eref{chsqfm}, we obtain
the results as displayed in Fig.~\ref{fig_datfit}. Just as expected,
the interference effect changes the shape of each resonance
significantly: the $\psi(4040)$ becomes narrower while the
$\psi(4160)$ and $\psi(4415)$ become wider than the previous
published results where the interferences between resonances are
neglected~\cite{sethr}.

The fit results are presented in Table~\ref{tab_datfit}, there are
four solutions found in the fit. It should be noted that the four
solutions have identical $\chi^2$, masses, and total widths for the
resonances, but different partial widths to lepton pairs. One
may suspect that the existence of four solutions is due to low
precision of the measurements. We will show in the next session
that multiple solution is a real effect. The improvement of the
precision of the measurements can not change the fact that there
are four solutions in fitting these data.

%\begin{table*}
\begin{table}
\caption{\label{tab_datfit} Four groups of solutions for the data
fitting. The four solutions have identical resonance masses ($M$)
and total widths ($\Gamma_t$), but significant different leptonic
partial widths ($\Gamma_{ee}$) and the relative phases ($\phi$).
%The fitted mass ($M$) and total width ($\Gamma_t$)
%for each resonance are the same for the four
%solutions.
The fit yields $\chi^2=91$, $A=(15.05\pm 0.59)$~nb, and
$B=(-1.64\pm 0.67)$~nb/GeV for all the solutions.
%The leptonic
%partial width ($\Gamma_{ee}$) and the relative phase ($\phi$) are
%different in different solutions.
}
\begin{center}
\begin{tabular}{cccc}
\hline\hline
  Parameter        & $\psi(4040)$ & $\psi(4160)$   & $\psi(4415)$ \\
\hline
%%%%%%%%%%%%%%%%%%%%%%%%%%%%%%%%%%%%%%%%%%%%%%%%%%%%%%%%%%%%%%%%%%%%%%%%%%%%
$M$ (MeV)       &$4034\pm  6$  &$4193\pm  7$  &$4412\pm 15$ \\
$\Gamma_t$ (MeV) &$  87\pm 11$  &$  79\pm 14$  &$ 118\pm 32$
\\\hline\hline
 $\Gamma^{(1)}_{ee}$ (keV)
           &~$0.66\pm 0.22$~&~$0.42\pm 0.16$~&~$0.45\pm 0.13$~ \\
$\phi^{(1)}$ (radian)
           & 0 (fixed)    &$2.7\pm 0.8$  &$2.0\pm 0.9$ \\\hline
$\Gamma^{(2)}_{ee}$ (keV)
           &$0.72\pm 0.24$&$0.73\pm 0.18$&$0.60\pm 0.25$ \\
$\phi^{(2)}$ (radian)
           & 0 (fixed)    &$3.1\pm 0.7$  &$1.4\pm 1.2$ \\\hline
$\Gamma^{(3)}_{ee}$ (keV)
           &$1.28\pm 0.45$&$0.62\pm 0.30$&$0.59\pm 0.20$ \\
$\phi^{(3)}$ (radian)
           & 0 (fixed)    &$3.7\pm 0.4$  &$3.8\pm 0.8$ \\\hline
$\Gamma^{(4)}_{ee}$ (keV)
           &$1.41\pm 0.12$&$1.10\pm 0.15$&$0.78\pm 0.17$ \\
$\phi^{(4)}$ (radian)
           & 0 (fixed)    &$4.1\pm 0.1$  &$3.2\pm 0.3$
           \\\hline\hline
\end{tabular}
\end{center}
\end{table}
%\end{table*}

\section{Multiple solutions}

The existence of four solutions in the fit described above can be
tested through the toy simulated data. The special
steps for this approach is:

\begin{enumerate}

\item The curve of the best fit in previous session (with
background contribution subtracted) is used as the probability
density function (PDF) to compute a set of experimental points at
different energies ($N_{pt}=100$ in this work) within the range
3.75~GeV $\leq \sqrt{s} \leq$ 4.85~GeV.

\item A relative error of 1\% and an absolute error of 0.01~nb are
added in quadrature and the total error is assigned to each data
point. The inclusion of a 0.01~nb absolute error is to weaken the
chisqure-weight of points with small cross sections.

\end{enumerate}

\begin{figure}[htbp]
\includegraphics[width=8.0cm]{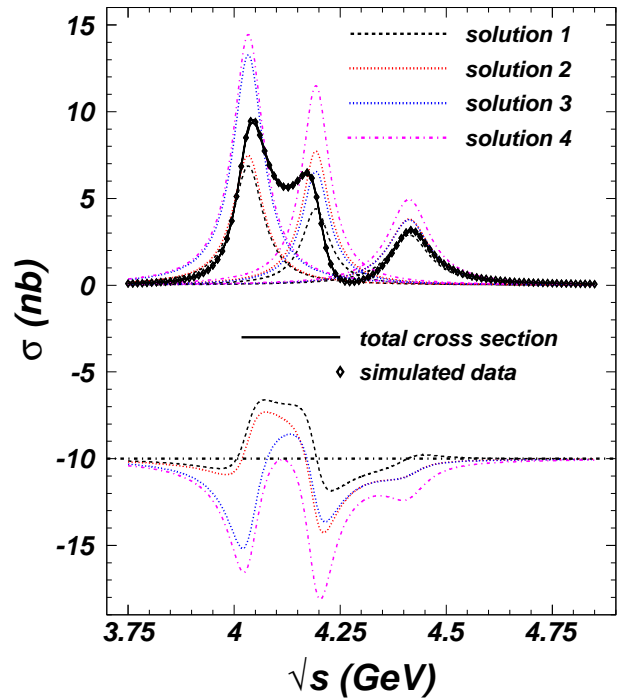}
\caption{\label{fig_shdt} Four groups of solutions obtained from
the fit to the generated data. For clearness, the interference
curves have been moved downward by 10~nb, the black dashed line at
$-$10~nb corresponds to zero cross section.}
\end{figure}

Fit the simulated data with the similar $\chi^2$ in
Eq.~\eref{chsqfm} but with only resonance cross section included
(i.e., $\sigma^{the.}(s) = \sigma^{res.}(s)$ in Eq.~\eref{sigfm}),
the four groups of solutions obtained are summarized in
Table~\ref{mtslndata} and displayed in Fig.~\ref{fig_shdt}.
Comparison of two tables obviously indicates that the central values of
the parameters are consistent with each other.

%\begin{table*}
\begin{table}
\caption{\label{mtslndata}Fit results for four groups of solutions
with the toy simulated data points. The definitions of
the parameters are the same as in Table~\ref{tab_datfit}. The fit
yields $\chi^2=1.0 \times 10^{-3}$ for all the solutions with the
expectation of $\chi^2=0$.}
\begin{center}
\begin{tabular}{cccc}
\hline\hline
  Parameter      & $\psi(4040)$  & $\psi(4160)$  & $\psi(4415)$ \\
\hline
$M$ (MeV)       &$4033.5\pm 0.3$ &$4192.8\pm 0.3$ &$4412.4\pm 0.4$  \\
$\Gamma_t$ (MeV) &$87.23\pm 0.49$ &$79.00\pm 0.53$ &$118.11\pm
0.56$
\\\hline\hline
 $\Gamma^{(1)}_{ee}$ (keV)
     &~$0.664\pm 0.005$~&~$0.417\pm 0.004$~&~$0.454\pm 0.003$~ \\
$\phi^{(1)}$ (radian)
               & 0 (fixed)      &$2.701\pm 0.012$&$2.002\pm 0.012$
               \\\hline
$\Gamma^{(2)}_{ee}$ (keV)
               &$0.723\pm 0.006$&$0.731\pm 0.005$&$0.596\pm 0.003$ \\
$\phi^{(2)}$ (radian)
               & 0 (fixed)        &$3.051\pm 0.001$&$1.432\pm 0.014$
               \\\hline
$\Gamma^{(3)}_{ee}$ (keV)
               &$1.283\pm 0.005$&$0.620\pm 0.006$&$0.590\pm 0.003$ \\
$\phi^{(3)}$ (radian)
               & 0 (fixed)        &$3.732\pm 0.006$&$3.789\pm 0.013$
               \\\hline
$\Gamma^{(4)}_{ee}$ (keV)
               &$1.397\pm 0.006$&$1.087\pm 0.008$&$0.774\pm 0.003$ \\
$\phi^{(4)}$ (radian)
               & 0 (fixed)        &$4.082\pm 0.005$&$3.218\pm 0.009$ \\
\hline\hline
\end{tabular}
\end{center}
\end{table}
%\end{table*}

From Table~\ref{mtslndata} and Fig.~\ref{fig_shdt}, we can see
that the largest $\Gamma_{ee}$ is more than twice the smallest
value in the four solutions. We also notice that the first
solution is identical to the BES published one~\cite{bespsires},
and the second solution is about the same as the one listed as the
best estimation of the partial widths of these states by the
PDG~\cite{PDG}.

%%It is really an important and hard work to choose from these
%%multiple solutions the physics one. If we follow the ``minimal
%%amplitude principle'' suggested in Ref.~\cite{Yuan:2009gd}, the
%%first solution listed in Table~\ref{mtslndata} will be selected as
%%the physics solution.

\section{Discussions}

As the $\Gamma_{ee}$ of the vector resonances are closely related
to the nature of these states, the choice among the distinctive
solutions affects the classification of the
charmonium and charmonium-like states observed in this energy
region.

The calculation of the $\EE$ partial widths of the $S$-wave
charmonium states is well summarized recently in Table~2 of
Ref.~\cite{wanggl}. We can see clearly that many of the
theoretical calculations give large $\psi(3S)$ and $\psi(4S)$
decay widths compared to the PDG values (about the same as the
second solution listed in Table~\ref{tab_datfit} above), but the
agreement with the third or the fourth solution is much better.

The $Y(4260)$ was proposed to be the $\psi(4S)$ state and the
$\psi(4415)$ be $\psi(5S)$ in Refs.~\cite{KTChao,Klempt}, we can
see that in this assignment, the calculated partial widths of
$\psi(4040)=\psi(3S)$ and $\psi(4415)=\psi(5S)$~\cite{KTChao}
agree well with the fourth solution listed in
Table~\ref{tab_datfit}.

Of course the possible mixing between $S$- and $D$-wave states
will change significantly the theoretical predictions of the
partial widths of these states~\cite{badalian}, and the QCD
correction, which is not well handled~\cite{wanggl}, may
also change the theoretical predictions significantly. So far, we
have no concrete criteria to choose any one of the solutions as
the physics one.

It should be noticed that if the $Y$ states are considered together
with the excited $\psi$ states in fitting the $R$-values, there
could be even more solutions, the situation may become more
complicated.

We also notice that the existence of the multiple solutions is due
to the inclusion of a free phase between two resonances, if these
phases can be determined by other means (either theoretically or
experimentally), it will be very helpful to know which solution
corresponds to the real physics.

\section{Summary}

Based on the $R$ scan data, the resonance parameters of excited
$\psi$-family resonances are fitted. We found that there are four
sets of solutions with exactly the same fit quality extracted from the
experimental data, but the leptonic partial widths among different
sets of solutions differentiate from each other significantly.
New information is needed to determine which
solution corresponds to the real physics.

\acknowledgments

This work is supported in part by the National Natural Science
Foundation of China (10775412, 10825524, 10935008), the Instrument
Developing Project of the Chinese Academy of Sciences (YZ200713),
Major State Basic Research Development Program (2009CB825203,
2009CB825206), and Knowledge Innovation Project of the Chinese
Academy of Sciences (KJCX2-YW-N29).

%%%%%%%%%%%%%%%%%%%%%%%%%%%%

\end{document}